# A O(|E|) Time Shortest Path Algorithm For Non-Negative Weighted Undirected Graphs

Muhammad Aasim Qureshi, Dr. Fadzil B. Hassan, Sohail Safdar, Rehan Akbar
Computer And Information Science Department
University Technologi PETRONAS
Perak, Malaysia

*Abstract*— **In most of the shortest path problems like vehicle routing problems and network routing problems, we only need an efficient path between two points—source and destination, and it is not necessary to calculate the shortest path from source to all other nodes. This paper concentrates on this very idea and presents an algorithms for calculating shortest path for (i) non-negative weighted undirected graphs (ii) unweighted undirected graphs. The algorithm completes its execution in O(|E|) for all graphs except few in which longer path (in terms of number of edges) from source to some node makes it best selection for that node.**

**The main advantage of the algorithms is its simplicity and it does not need complex data structures for implementations.**

I. INTRODUCTION

Shortest Path Problem can formally be defined as follows:

Let G be a graph such that G = (V, E), where V = { v1, v2, v3, v4, …, vn } and E = { e1, e2, e3, e4, …, em } such that |V| = n and |E| = m. G is an undirected weighted connected graph having no negative weight edge, with pre-specified source vertex 's' and destination vertex 't' such that s ∈ V and d ∈ V. We have to find simple path from s to t with minimum most total edge weight.

Theoretical Computer Science is one of the most important and hardest areas of Computer science (TCS) [17][18][19]. The single-source shortest paths problem (SSSP) is one of the classic problems in algorithmic graph theory of TCS. Since 1959, all theoretical developments in SSSP for general directed and undirected graphs have been based on Dijkstra's algorithm, visiting the vertices in order of increasing distance from **s**. As a matter of fact many real life problems can be represented as SSSP. As such, SSSP has been extensively applied in communication, computer systems, transportation networks and many other practical problems [1].

The complexity of Dijkstra's algorithm [10] has been determined as $O(n^2 + m)$ if linear search is used to calculate the minimum [2]. A new heap data structure was introduced by [4][5] to calculate the minimum which resulted he complexity to O(m log n). The complexity was further improved [9] when Fredman and Tarjan developed Fibonaccii heap. The work in [9] was an optimal implementation of Dijkstra's algorithm in a comparison model since Dijkstra's algorithm visits the vertices in sorted order. Using fusion trees of [8], we get an O(m (log n) ½ ) randomized bound. Their later atomic heaps give an O(m + n log n/log log n) bound presented in [7]. Afterwards, in [11][12][16] priority queues gave an O(m log log n) bound and an O(m + n(log n1+ε)½) bound. These bounds are randomized assuming that we want linear space. Afterwards [14] reduced it to O(m + n(log n log log n) ½) and next year [15] improved it with randomized bound to O(m + n(log n1+ε) 1/3) .

Priority queue presented in [6] for SSSP improved the shortest path cost giving a running time of O(m + n(log C) ½) where C was the cost of the heaviest edge. Next work by [13] to reduced the complexity to O(m + n (3 log C log log C) 1/3 ) expected time and [15] presented a further improvement to O(m + n(log C) 1/4+ε). [3] presented an algorithm and claimed that it will out class dijekstra's algorithm.

Contrary to Dijekstra and many others this algorithm attacks the problem from both ends—source and destination. It searches source node (i.e.'s'), starting from destination node (i.e. 't') and on the other side searches destination node starting from source node in parallel.

II. BASIC IDEA

This algorithm is basically an extension of the work done in [20]. The basic idea can be best described using an analogy of two distinct persons involved in a task of searching a path between starting point (Point1) and finish point (Point2) of a labyrinth. First person, A, starts from point1 and second person starts from point2 as illustrated in fig. 1. A explores all possible paths searching for either B or point2. and in the same way second man, B, starts exploring all the paths starting from point2 looking for point1 or A as illustrated in fig. 2. They meets on the their way (see fig. 3) to their destination and as soon as they meet they exchange and combine their information about the path they have traversed and can easily be made a path along with total cost of the path.

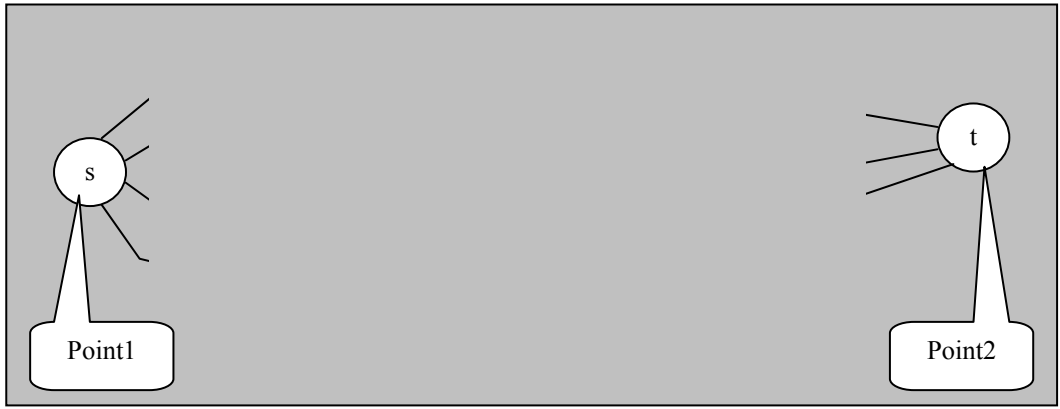

Figure 1:(step 1)Person A starts from point1 and Person B starts from pointB

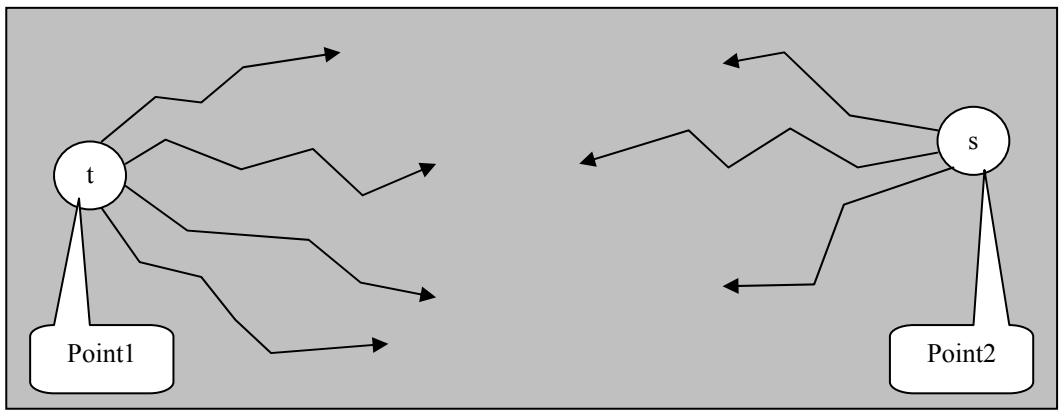

Figure 2: at next levels Both A and B are exploring different paths in search of one another

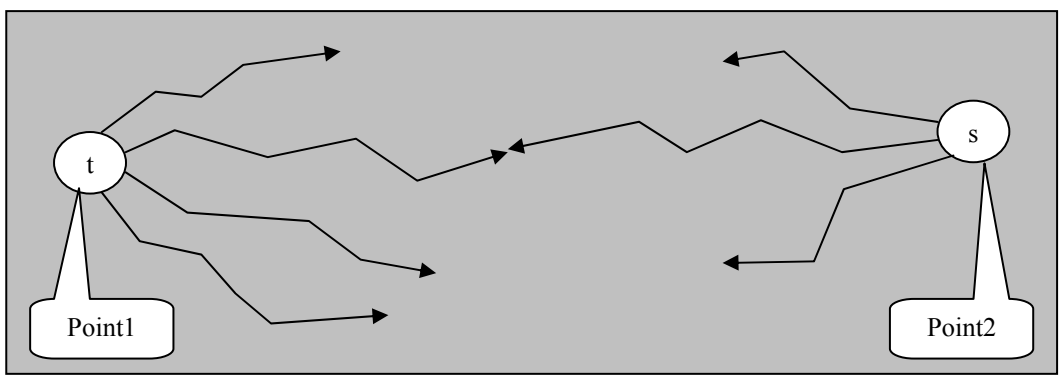

Figure 3: Both A and B meet at some point and by interchanging information can make the whole path

## III. ALGORITHM

### A. Algorithm Input Constraints

This Algorithms runs well for all graphs but for the few having following properties:

Len(Pi(s,w)) > Len(Pj(s,w))

$$\underset{\text{for } Pi}{\overset{k-1}{\underset{i=0}{\sum}} w(x_i, x_{i+1})} \angle \underset{\text{for } Pj}{\overset{l-1}{\underset{i=0}{\sum}} w(y_i, y_{i+1})}$$

where k ≠ l and x0 = y0 = s and xk = yk = w and Pi and Pj are paths from s to w such that

Pi = { x0 ,x1 ,x2 , . . . . . . . . ,xk } and

Pj = { y0 ,y1 ,y2 , . . . . . . . . ,yl }

### B. Algorithm Definition

This algorithm has three main parts namely, PartA, PartB and PartC. Part A and PartB are identical. Both are searching the footmarks of the other Part. PartA is concerned with the search for the shortest path from the source node **s** to the destination node **t** and PartB targeting **s,** starts its search from **t**. Both PartA and PartB are replicas of one another and perform similar actions. The two parts are running in pseudo-parallel fashion, exploring nodes of the graph level by level.

First of all data structures are initialized as:

$$\underset{U \in V(G)/s/t}{\forall} \left[ \pi_u \leftarrow \text{NIL}, \text{CST}_u \leftarrow -\infty, \text{CLR}_u \leftarrow \textbf{YELLOW}, \text{DST}_u \leftarrow -1 \right]$$

**s** and **t** is initialized with NIL, 0, GRAY, 0 respectively.

Each part—PartA and PartB, starts investigation from its respective starting node (**s** for PartA and **t** for PartB) and explores all its neighboring nodes. So let's say **s** and **t** are level-0 nodes (**NL0**) and all nodes being explored from these nodes will be level-1 nodes (**NL1**) and all the nodes explored from level-1 node are level-2 (**NL2**) and so forth.

Other than NL0 all nodes have to calculate and keep track of the best cost (i.e. $CST_x$) along with the parent node ($\Pi_x$) making the best cost. The track of the status of each node is kept by the coloring them with specific colors (i.e. CLR). Details are as below:

GREEN: the node is neither explored nor traversed. It means that algorithm has not yet come across this node

YELLOW: node is explored by some node. This node can still be explored by other node(s).

RED: the node is explored and traversed. Each YELLOW node is picked and its neighbors are explored and then it is painted as RED.

While exploring, neighboring nodes from any node (say **p**), it calculates the cost of the node being explored (cost of the path starting from **NL0** point to the node in hand i.e. **h** ) in order to calculate the best cost so far (i.e. $CST_h$) and best parent of the node making its cost minimum (i.e. $\Pi_h$). $CST_h$ and $\Pi_x$ will be calculated as below:

Cost at node **h** is

old_$CST_h$ = $CST_h$

$CST_h$ = min($CST_h$ , $CST_p$ + $e_{p,h}$)  (1)

and

if $CST_h$ = $CST_p$ + $e_{p,h}$

then  $\Pi_h$ ← **p**  (2)

Initially all nodes are painted as GREEN (while initializing except source and destination nodes) and as soon as a node is explored during the traversal it is painted as YELLOW and as soon as any node completes its traversal (i.e. all its neighbors are explored i.e. painted YELLOW) it is painted RED.

Until all the YELLOW nodes are converted to RED of some level no node is selected for traversal from the next level. As soon as one level is completed the control is switched to the other part of the algorithm to proceed and it also performs the same steps.

During these traversals if some node is found that was marked RED by other part then two nodes **p** and **h** are stored along with the total cost the complete path calculated as

old_SPCST ← SPCST

SPCST ← min ( SPCST, $CST_h$ + $CST_p$ + $e_{p,h}$)  (3)

If SPCST = $CST_h$ + $CST_p$ + $e_{p,h}$

Then SP ← ph  (4)

In this way all possible paths are covered and their costs are stored. This algorithm continues until there is any YELLOW or RED node in the graph.

When all nodes are colored RED, PartA and PartB of algorithm stops and PartC is invoked. PartC using a simple linear search algorithm searches for the path with minimum cost from the stored costs using the nodes that were stored.

### C. Working Example of the Algorithm

NOTE: For this example the color scheme is changed to WHITE, GRAY and BLACK to get better display.

The algorithm starts with PartA (i.e. from source **s**) and marking the cost of the node as 0 and painting it GRAY as shown in fig. 4.

On the other end PartB starts in parallel from **t** and marking its cost as 0 and painting it GRAY as shown in fig. 4.

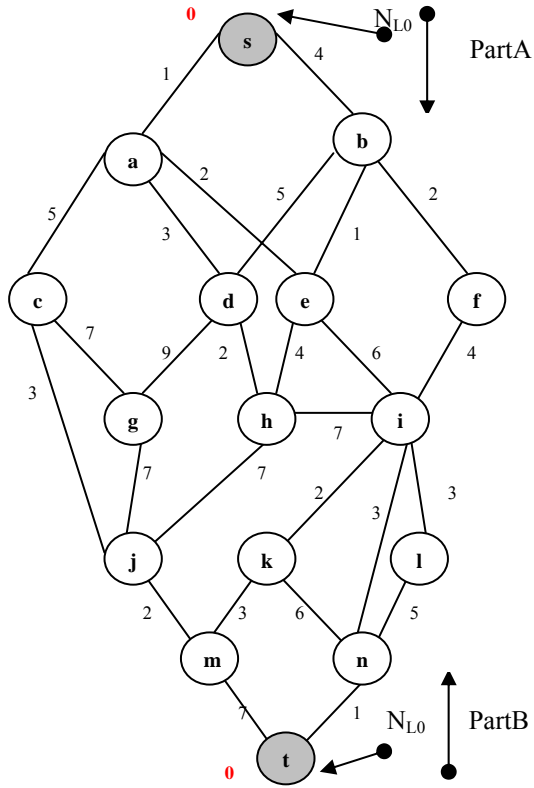

Figure 4: Starting PartA and PartB from s and t respectively

Continuing from **NL0**, the algorithm starts with investigation from p = 's' exploring all it neighbors, one by one, randomly. Assume **h** = '**b**' is picked and now its cost and parent is adjusted using (1), (2) (its cost is 4 as shown in red color i.e. 0+4) and marking **p** i.e. '**s**' as its parent being shown as green line and marking it explored by painting it GRAY. Then next neighbor is chosen and the whole process is repeated and then next neighbor is picked. This process continues until all the neighbors are painted GRAY. Upon the completion of the exploration process **p** is painted BLACK.(see fig. 5)

On the other end PartB starts its processing from NL0, and picks **p** = '**t**' for traversal. All its neighbors are explored, one by one, randomly. Supposing **h** = '**n**' is picked and now its cost and parent is adjusted using (1), (2) (its cost is 1 as shown in red color). **t** is marked as the parent of **n**, shown with blue line and its status is changed to explored by painting it GRAY. Then next neighbor is chosen and the whole process is continued until all the neighbors are painted GRAY. Upon the completion of the traversal process **p** is painted BLACK.(see figure 5)

Now investigating NL1 nodes (a and b) one by one and checking their neighbors and performing actions like marking and/or adjusting costs and parents (using (1) and (2)) and painting neighbors GRAY. All NL1 nodes are painted BLACK (see fig. 6) one by one.

Same process is being repeated in PartB on NL1 nodes (m and n).(see fig. 6)

PartA is repeating same steps that were performed previously but now on NL2 nodes(c, d, e, and f) (see fig. 7)

In partB doing the same steps as in PartA e.g. exploring all nodes of NL2 (I, j, k, and l) one by one, the notable point, here, is that node **i** explores **e** and **f** and finds them already traversed

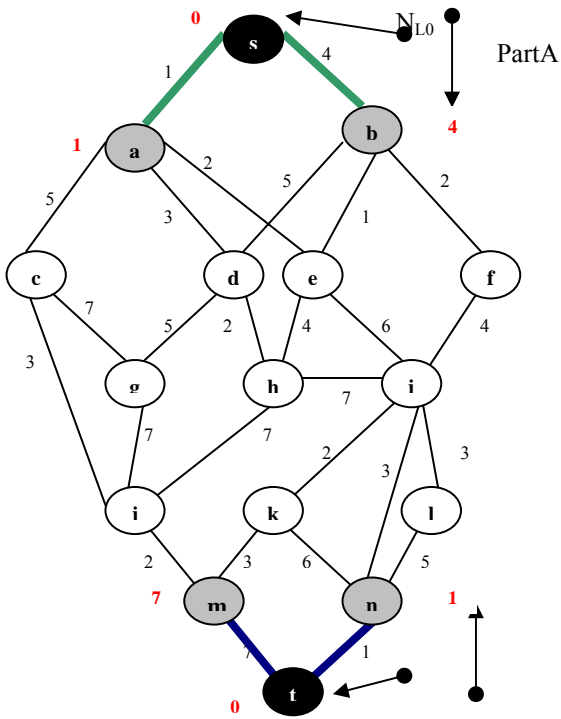

Figure 5: Traversing level-0 nodes from both sides

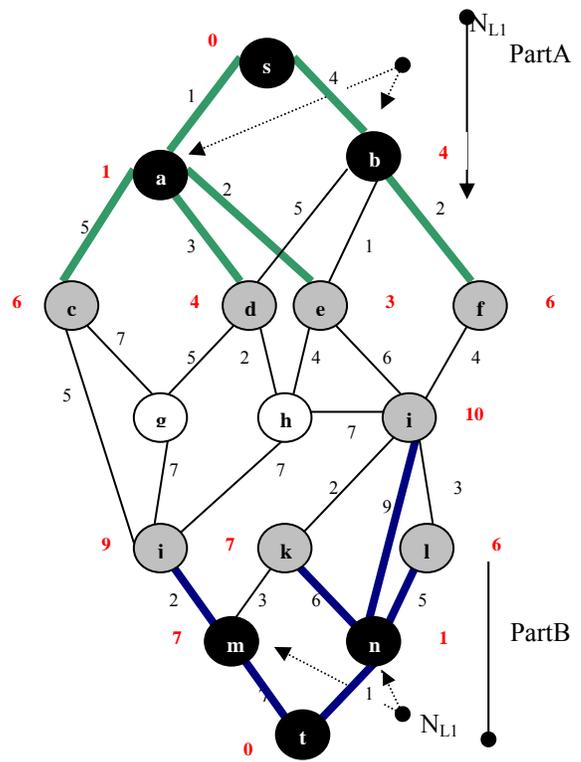

Figure 6: Traversing Level-1 nodes from both sides

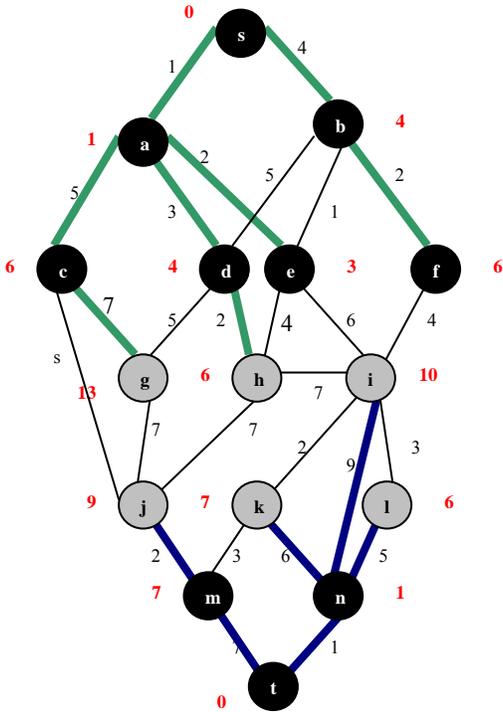

Figure 7: Investigating nodes (c,d,e,f) (only processing of PartA

(painted BLACK) by other part (i.e. PartA) so algorithm here stores 'I' and 'e' and the cost 3+6+9 (using (3) and (4)) and then stores I and f and cost 6+4+9 (using (3) and (4)). This path is marked with yellow line. Here we explored and stored three steps(see fig. 8)

PartA starts exploring NL3 nodes (i.e. g and h) that are GRAY. Performing the same steps PartB did in the previous step (see fig. 8). Here three new paths are explored and stored along with their costs.

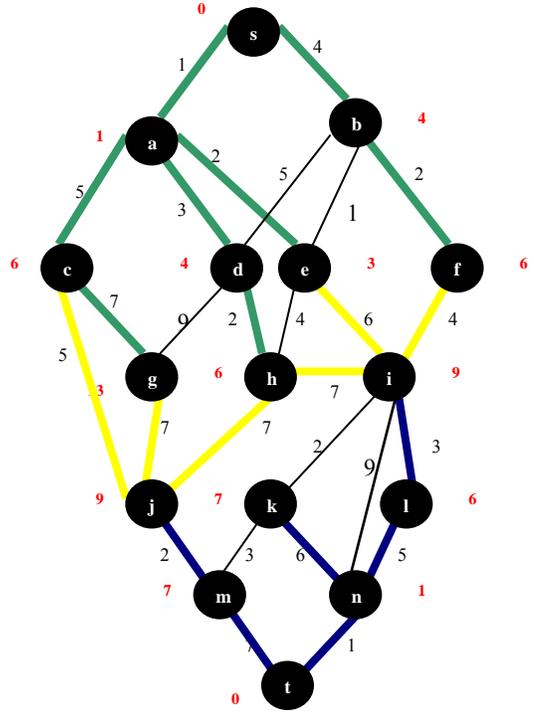

Figure 9: Collision of two Parts of algorithm making SP

PartB has no GREY nodes to continue its traversal. So it will terminate. As all the nodes in the graph are now BLACK so PartA and PartB terminates.

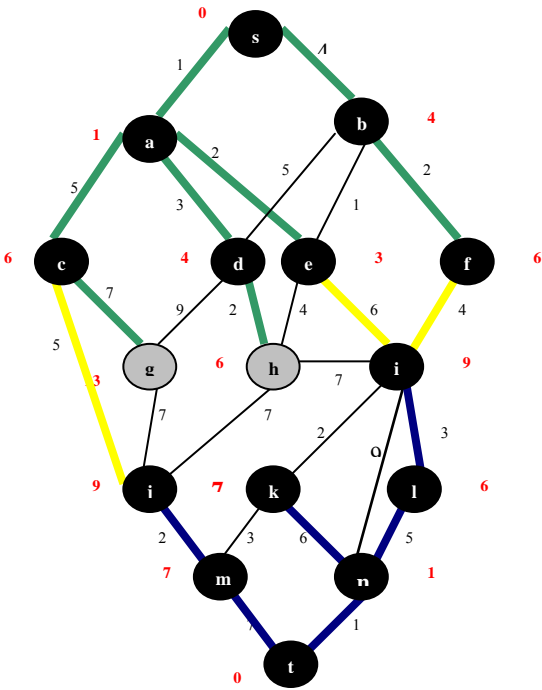

Figure 8: Traversing i, j, k, l

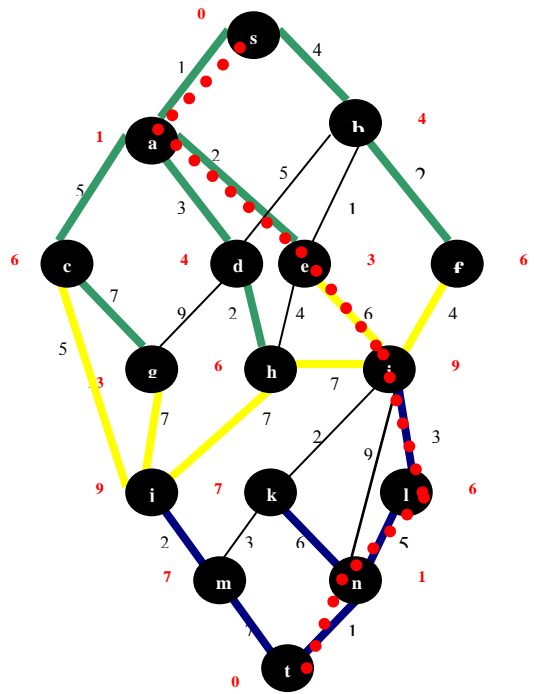

Figure 10: linear search in calculated paths resulted in SP

PartC invokes and calculates the minimum out of all costs calculated so far and determines the shortest path (see fig. 10) (in the algorithm you will find it embedded in the PartA and PartB calculations)

*D. Pseudo Code*

In this pseudo-code we are using four subroutines—Shortest_Path_Algorithm, Initialize, PartA_B_C, Print_Path. Shortest_Path_Algorithm is the main subroutine that invokesother subroutines. Initialize is for all kind of initializations required for the execution of the algorithm. Then PartA_B_C is invoked twice with different queues making it PartA and PartB. PartC is embedded in the PartA_B_C in the end. Finally Print_Path is invoked to print the shortest path.

*1) Legend being used in the algorithm:*

CLR: Color – can be one of the three—GREEN (no processing has yetstarted on this node) , YELLOW ( processing has started on this node) and RED ( Processing on the current node has completed

$CLR_v$: Color of v

$CLR_u$: Color of u

$CST_v$: Cost of v (i.e. minimum cost from source to v)

$DST_v$: Number of Edges in the path from source s to current node v

$RED_o$: Color RED painted by other part of the algorithm e.g. if currently PartA is being executed then it will be referring to a node that would be painted by PartB

$RED_t$: Color RED painted by this part of the algorithm e.g. if currently PartA is being executed then it will be referring to a node that would be painted by PartA

YELLOW_: node is marked YELLOW and it is inserted in the next queue and should not be processed from current queue

$Q_s$: Queue that is being used by PartA

$Q_s$: Queue that is being used by PartB

SPCST: shortest Path Cost

SP: Shortest Path

**Shortest_Path_Algorithm ()**
```
(1) .. Initialize ()
(2) .. while (Qs ≠ ∅ AND Qt ≠ ∅ )
(3) .. do ..
(4) .. .. PartA_B_C(Qs)
(5) .. .. PartA_B_C(Qt)
(6) .. Print_Path(SP)
```
-------------------------------------

**Initialize ()**
```
(1) .. for each v ∈ V
(2) .. do ..
(3) .. .. CLRv ← GREEN
(4) .. .. Πv ← ∅
(5) .. .. CSTv ← −∞
(6) .. .. DSTv ← −∞
..
(7) .. CLRs = YELLOW
(8) .. CLRt = YELLOW

(9) .. ENQueue (Qs, s)
(10). EnQueue (Qt, t)
```
-------------------------------------

**PartA_B_C(Q)**
```
(1) Qtmp ← ∅
(2) while Q ≠ ∅
(3) .. then u ← DeQueue(Q)
(4) .. .. if u ≠ YELLOW_
(5) .. .. .. Then for each v ∈ Adj[u]
(6) .. .. .. do if CLRv = GREEN
(7) .. .. .. .. then CLRv ← YELLOW
(8) .. .. .. .. .. EnQueue (Qtmp, v)
(9) .. .. .. .. .. Πv ← u
(10). .. .. .. .. CSTv ← CSTu + eu,v
(11). .. .. .. .. DSTv ← DSTu + 1
(12). .. .. .. Else if CLRv = YELLOW
(13). .. .. .. .. Then if CSTv > CSTu + eu,v
(14). .. .. .. .. .. Then If DSTv=DSTu & CLRv≠YELLOW_
(15). .. .. .. .. .. .. Then EnQueue (Qtmp, v)
(16). .. .. .. .. .. .. CLRv ← YELLOW_
(17). .. .. .. .. .. Πv ← u
(18). .. .. .. .. .. CSTv ← CSTu + eu,v
(19). .. .. .. .. .. DSTv ← DSTu + 1
(20). .. .. .. Else if CLRv = REDt
(21). .. .. .. .. Then  if CSTv > CSTu + eu,v
(22). .. .. .. .. .. Then print "wrong graph"
(23). .. .. .. .. .. Terminate Algorithm
(24). .. .. .. Else if CLRv = REDo
(25). .. .. .. .. Then
(26). .. .. .. .. .. Πv ← u
(27). .. .. .. .. .. If CSTu +CSTv +eu,v<SPCST
(28). .. .. .. .. .. Then .
(29). .. .. .. .. .. .. SP ← "uv"
(30). .. .. .. .. .. .. SPCST←CSTu +CSTv +eu,v
```
-------------------------------------

**Print_Path(SP)**
```
(1) .. PTH[1 to DSTv + DSTu + 1]

(2) .. i←DSTSP[1]
(3) .. PTH[i] ← u
(4) .. While p is not eual to NULL
(5) .. Do
(6) .. .. i ← i - 1
(7) .. .. PTH[i] ← p ← Πp

(8) .. i←DSTSP[1] + 1
(9) .. PTH[i] ← v
(10). While p is not eual to NULL
(11). Do
(12). .. i ← i + 1
(13). .. PTH[i] ← p ← Πp
(14).
(15). For i←1 to DSTv + DSTu + 1
(16). Do
(17). .. Print PTH[i], ","
```

*E. Complexity*

This example shows that the algorithm successfully completes its execution for targeted graphs. The algorithm starts with two parts both traversing and covering the neighbors using edges. And a node never re-covers the node that it has already covered. Both parts are moving at the same pace (i.e. covering nodes level by level) so both parts will be covering almost same number of nodes (on average). So in this way each part will be covering E/2 edges making total = E.

Embedded in Part A and PartB, PartC calculates shortest path using the technique of linear search. As there can not be more than E paths (in the worst case) so linear search can take maximum E time to complete its execution and find out minimum cost path. so it make total complexity to E+E=2E which is O(E).

The main advantage as well as the beauty of this algorithm is that it is very simple, easy to learn and easy to implement. At the same time it does not require complex data structures.

So this algorithm can be applied for problems like vehicle routing, where the maps of the roads grow always in hierarchical fashion and very rarely a situation occur in which a long path give a smaller cost.

IV. SAME ALGORITHM FOR DIFERENT TYPES OF GRAPHS

Applying this algorithm on weighted directed graphs, it produced a quick result as it solves the given problem from two ends (i.e. source and destination).

Minor modification is required to calculate the shortest path for unweighted directed/undirected graphs of all types without any bound and/or condition. Modification that is required is to terminate the algorithm as soon as one investigating node checks some node that is colored GRAY by other part of the algorithm. In other words we can say that as soon as two parts collide for the first time. Algorithm is terminated and combining the paths of two nodes will give the shortest path. Though this algorithm also work in O(E) in worst case that is also the complexity of BFS but results showed that it conclude quite efficiently and calculates the path in less time.

V. CONCLUSION

This algorithm is very efficient and robust for the targeted graphs due to its simplicity and along with it the constant factor is quite negligible. For all kinds of unweighted graphs, algorithm showed promising results. Though, it does not improve asymptotic time complexity but in terms of he number of processing steps its results were much better (most of the times) than Breadth First Search. In nonnegative weighted undirected graphs (except few) this is very fast and efficiently convergent algorithm for targeted graphs.